\documentclass[11pt,epsf,letterpaper]{article}%
\usepackage{color}
\usepackage{amsmath}
\usepackage{amsfonts}
\usepackage{verbatim}
\usepackage{amssymb}
\usepackage{graphicx}
\usepackage{epstopdf}
\usepackage{mathrsfs}%
\providecommand{\U}[1]{\protect\rule{.1in}{.1in}}
\begin{document}

\title{Exact hairy black brane solutions in $AdS_{5}$ and holographic RG flows}
\author{$^{(1)}$Andr\'{e}s Ace\~{n}a, $^{(2)}$Andr\'{e}s Anabal\'{o}n, and $^{(3)}%
$Dumitru Astefanesei\\\textit{$^{(1)}$Instituto de Ciencias B\'asicas, Universidad Nacional de Cuyo,
Mendoza, Argentina.} \\\textit{$^{(2)}$Departamento de Ciencias, Facultad de Artes Liberales y}\\\textit{Facultad de Ingenier\'{\i}a y Ciencias, Universidad Adolfo
Ib\'{a}\~{n}ez, Vi\~{n}a del Mar.} \\\textit{$^{(3)}$ Instituto de F\'\i sica, Pontificia Universidad Cat\'olica de
Valpara\'\i so,} \\\textit{ Casilla 4059, Valpara\'{\i}so, Chile} \\acena.andres@gmail.com, andres.anabalon@uai.cl, dumitru.astefanesei@ucv.cl}
\maketitle

\begin{abstract}
We construct a general class of \textit{exact regular} black hole solutions
with toroidal horizon topology in $5$-dimensional Anti-deSitter gravity with a
self-interacting scalar field. With these boundary conditions and due to the 
non-trivial backreaction of the scalar field, the no-hair theorems can be 
evaded so that an event horizon can be formed. The scalar field is regular 
everywhere outside the curvature
singularity and it vanishes at the boundary where the potential is finite. We
study the properties of these black holes in the context of AdS/CFT duality
and comment on the dual operators, which saturate the unitarity bound. We
present exact expressions for the beta-function and construct a c-function
that characterizes the RG flow.

\end{abstract}

\section{Introduction.}

Motivated by AdS/CFT duality \cite{Maldacena:1997re}, there has been extensive
work constructing solutions in Anti de Sitter (AdS) spacetime. The finite
temperature dynamics of holographic field theories can be related to the
thermodynamics of black holes on the gravity side. Of particular interest are
black hole solutions with scalar hair (see, e.g., \cite{Giusto:2012gt,
Gubser:2008px, Hartnoll:2008vx, Correa:2010hf, Basu:2010uz, Anabalon:2012sn,
Brihaye:2012cb, Chowdhury:2012ff, Bardoux:2012tr} for some recent
applications). However, many of these solutions (for non-trivial scalar
potentials) were generated numerically only\footnote{In fact, in many relevant
examples for holography, by imposing constraints on the form of the scalar
potential just the asymptotic and near horizon expansions of the metric are
presented.} and so the physics of these solutions in the context of holography
can be just partially investigated.

In this letter, we present a class of \textit{exact}, toroidal black hole
solutions in AdS$_{5}$ gravity with a self-interacting scalar field --- the
non-trivial backreaction of the scalar field can produce an event horizon. To
the best of our knowledge, this is the first example of a $5$-dimensional
exact, regular, neutral, hairy black hole solution. These solutions are
important for understanding the no-hair theorems when the horizon does not
have spherical topology (see, e.g., \cite{Hertog:2006rr}) and are constructed
by using the method of \cite{Anabalon:2012ta} (see, also,
\cite{Anabalon:2012ih}).

Within AdS/CFT duality, the physics of this family of solutions is very rich.
The classical (super)gravity regime corresponds to the large-N, strong 't
Hooft coupling limit of the gauge theory. Using exact AdS gravity
backgrounds\footnote{By including additional background fields, which maintain
the asymptotic boundary behaviour, the conformal symmetry can be broken.}, it
is possible to investigate in detail theories that have some similar features
with four-dimensional QCD, e.g. confinement/deconfinement transitions
\cite{Witten:1998qj}, non-trivial speed of sounds, bulk viscosities
\cite{gubser}, RG flows ($\beta$-functions) \cite{Akhmedov:1998vf, verlinde,
Freedman:1999gp}, hydrodynamic properties \cite{Bhattacharyya:2008jc}, et
cetera. The existence of exact hairy black branes in AdS opens the exciting
possibility of investigating the strong regime of certain field theories in a
well controlled setting.

Interestingly enough, a part of the scalar potential we consider is controlled
by a tunable parameter, $\alpha$, so that we can obtain exact domain wall
solutions when $\alpha=0$ and exact planar black holes when $\alpha\neq0$.
This allows to explore how the extended literature on domain walls and fake
supergravity potentials apply to this case. Since the black hole solutions are
analytic, we obtain exact beta-functions. We would like to point out that the
expressions for the domain wall and black hole beta-functions coincide, though
the physics is obviously different.

For tachyonic scalars, it was shown in \cite{Henneaux:2006hk} that the slow
$a$-branch\footnote{The asymptotic form of the scalar field is $\phi=\frac
{a}{r^{\Delta_{-}}}+\frac{b}{r^{\Delta_{+}}},$ where $\Delta_{\pm}=2\left(
1\pm\sqrt{1+\frac{l^{2}m^{2}}{4}}\right)  $ and $l$ is the AdS radius. The
slow branch is the one controlled by $\Delta_{-}$.} contributes to the
boundary charges. The mass was chosen to be $m^{2}=-3l^{-2}$,
where $l$ is the $AdS_{5}$ radius, which is
indeed above the Breitenlohner-Freedman bound and it belongs to the mass
spectrum of the lowest short multiplet of $PSU(2,2|4).$ This implies
that the linearized spectrum of this model coincides with the one of type
$IIB$ supergravity compactified on $S^{5}$. Moreover, the mass
saturates the unitarity bound, in \cite{Ishibashi:2004, Henneaux:2006hk}, it
was argued that, when the unitarity bound is violated $\Delta_{-}\leq1$, the
Klein-Gordon inner product is divergent when evaluated for these
configurations. Indeed, the Klein Gordon inner product is well defined only
for solutions of the \textit{linearized} Klein-Gordon field equation, namely
perturbations over the configurations that we are considering here. These
linearized perturbation seems to be normalizable only if Dirichlet or, as is
the case of this work, AdS invariant boundary conditions are imposed on them
\cite{Andrade:2011dg}.

We will present the thermodynamical properties of our solutions and also
discuss in detail some of their interesting `holographic' features.

\section{The setup}

The action we are interested in is%

\begin{equation}
I[g_{\mu\nu},\phi]=\frac{1}{2\kappa}\int_{M}d^{5}x\sqrt{-g}\left[  R-\frac
{1}{2}\partial_{\mu}\phi\partial^{\mu}\phi-V(\phi)\right]  -\frac{1}{\kappa
}\int_{\partial M}d^{4}x\sqrt{-\gamma}K
\end{equation}
where $V(\phi)$ is the scalar potential and we use the convention $\kappa=8\pi
G_{N}$. Here, $K$ is the trace of the extrinsic curvature of the boundary
$\partial M$ as embedded in $M$ and $\gamma_{ab}$ is the induced metric on the
boundary. Since we set $c=1=\hbar$, $\left[  \kappa\right]  =M_{P}^{-3}$ where
$M_{P}$ is the five-dimensional Planck scale.

The equations of motion for the metric and dilaton are
\begin{equation}
E_{\mu\nu}=R_{\mu\nu}-\frac{1}{2}g_{\mu\nu}R-\frac{1}{2}T_{\mu\nu}^{\phi
}=0\,\,,\,\,\,\,\,\,\,\,\frac{1}{\sqrt{-g}}\partial_{\mu}\left(  \sqrt
{-g}g^{\mu\nu}\partial_{\nu}\phi\right)  -\frac{\partial V}{\partial\phi}=0
\end{equation}
where the stress tensors of the matter fields is
\begin{equation}
T_{\mu\nu}^{\phi}=\partial_{\mu}\phi\partial_{\nu}\phi-g_{\mu\nu}\left[
\frac{1}{2}\left(  \partial\phi\right)  ^{2}+V(\phi)\right]  \,\,\,\,
\end{equation}

For simplifying our analysis, we use the following ansatz for the metric:
\begin{equation}
ds^{2}=\Omega(x)\left[  -f(x)dt^{2}+\frac{\eta^{2}dx^{2}}{f(x)}+d\Sigma
^{2}\right]  \label{Ansatz}%
\end{equation}
where the parameter $\eta$ was introduced to obtain a dimensionless radial
coordinate $x$, $\Omega(x)$ is the conformal factor, and $d\Sigma^{2}%
=\sum\limits_{a=1}^{3}dx_{a}^{2}$.

There are three independent (combinations of) equations of motion. Two of them are%

\begin{align}
E_{t}^{t}-E_{r}^{r}  &  =0\Longrightarrow\phi^{\prime2}=\frac{9\left(
\Omega^{\prime}\right)  ^{2}-6\Omega\Omega^{\prime\prime}}{2\Omega^{2}%
}\label{eqphi}\\
E_{t}^{t}-E_{a}^{a}  &  =0\Longrightarrow2\Omega f^{\prime\prime}%
+3\Omega^{\prime}f^{\prime}=0
\end{align}
where the derivatives are with respect to $x$. The remaining equation is more
complicated, but it is worth emphasizing that the potential can be obtained by
solving it.

\section{Solutions}

As in \cite{Anabalon:2012ta}, we choose the conformal factor
\begin{equation}
\Omega(x)=\frac{\nu^{2}x^{\nu-1}}{\eta^{2}\left(  x^{\nu}-1\right)  ^{2}}%
\end{equation}
so that we can solve \eqref{eqphi} to get
\begin{equation}
\phi=l_{\nu}^{-1}\ln(x)\,\,\,\,\,\,\,\,\,,\,\,\,\,\,\,l_{\nu}^{-1}=\sqrt
{\frac{3(\nu^{2}-1)}{2}}%
\end{equation}
where $\nu^{2}>1$. A canonically normalized scalar field has the dimension
$[l_{\nu}]=[L^{\frac{3}{2}}]$, but with our normalization the scalar field is
dimensionless. The parameter $\nu$ labels different hairy solutions.

The other metric function is
\begin{equation}
f(x)=-\frac{\Lambda}{6}+\alpha\left[  \frac{4}{3\left(  \nu^{2}-25\right)
\left(  9\nu^{2}-25\right)  }+\frac{x^{\frac{5}{2}}}{12\nu^{3}}\left(
\frac{x^{\frac{3\nu}{2}}}{3(3\nu+5)}+\frac{x^{-\frac{3\nu}{2}}}{3(3\nu
-5)}-\frac{x^{\frac{\nu}{2}}}{(\nu+5)}-\frac{x^{-\frac{\nu}{2}}}{(\nu
-5)}\right)  \right]  \label{f}%
\end{equation}
and the dilaton potential is
\begin{align}
V(\phi)  &  =\frac{(9\nu^{2}-25)e^{-\phi l_{\nu}}}{4\nu^{2}}\left[
\Lambda-\frac{8\alpha}{\left(  \nu^{2}-25\right)  \left(  9\nu^{2}-25\right)
}\right]  \left[  \frac{\left(  \nu+1\right)  }{2\left(  3\nu-5\right)
}e^{\nu\phi l_{\nu}}+\frac{\left(  \nu-1\right)  }{2\left(  3\nu+5\right)
}e^{-\nu\phi l_{\nu}}+\frac{5\left(  \nu^{2}-1\right)  }{(9\nu^{2}-25)}\right]
\nonumber\\
&  +\frac{\alpha e^{\frac{3}{2}\phi l_{\nu}}}{2\nu^{3}}\left[  \frac{5\left(
\nu^{2}-1\right)  }{\nu^{2}-25}\left(  \frac{e^{\frac{\nu}{2}\phi l_{\nu}}%
}{3\nu+5}+\frac{e^{-\frac{\nu}{2}\phi l_{\nu}}}{3\nu-5}\right)  +\frac{1}%
{3}\left(  \frac{\left(  \nu-1\right)  e^{\frac{3\nu}{2}\phi l_{\nu}}}{\left(
\nu+5\right)  \left(  3\nu+5\right)  }+\frac{\left(  \nu+1\right)
e^{-\frac{3\nu}{2}\phi l_{\nu}}}{\left(  \nu-5\right)  \left(  3\nu-5\right)
}\right)  \right]  \label{Pot}%
\end{align}
Note that the potential and the configuration are invariant under the change
of $\nu\rightarrow-\nu$. At this point we observe that the potential has two
parts, one which is controlled by the cosmological constant $\Lambda=-\frac
{6}{l^{2}}$ and the other by $\alpha$ that is an arbitrary parameter with the
same dimension as $\Lambda$, namely $[\alpha]=[L^{-2}]$. The rank of the
coordinate $x$ can be taken to be either $x\in(0,1]$ or $x\in\lbrack1,\infty
)$. The scalar field is negative in the first case, but positive in the other
range. Since the dilaton potential has no obvious symmetry, this allows to
cover it completely. The singularities of the metric (and the scalar field)
are at $x=0$ and $x=\infty$, but they are enclosed by an event horizon. It is
easy to see that, for any value of $\Lambda$, $\nu$, and $x_{+}$, there is an
$\alpha$ such that $f(x_{+})=0$. Indeed, this simply follows from the fact
that $f(x)$ is linear in $\alpha$. Therefore, there are black holes in an open
set of the parameter space. Moreover, a simple inspection of the function
(\ref{f}) shows that is regular for every positive $x\neq0$ and $x\neq\infty$.

At the boundary, $x=1$, the scalar field vanishes, the potential is
$V=2\Lambda$ and, moreover, $\left.  \frac{dV}{d\phi}\right\vert _{\phi=0}=0$,
$\left.  \frac{d^{2}V}{d\phi^{2}}\right\vert _{\phi=0}=-\frac{3}{l^{2}}$.
Therefore, the scalar mass is indeed above the Breitenlohner-Freedman
bound and it matches the mass of some scalars in the spectrum of the lowest short multiplet of
$PSU(2,2|4)$. A numerical exploration of other extrema seems to indicate
that the only one is this maximum at the origin of the potential.

As is well known, the only relevant energy condition in AdS spacetimes
is the null energy condition, which is trivially satisfied for hairy solutions
of the form (\ref{Ansatz}). In an orthonormal frame $e_{\mu}^{a}$,
the energy momentum tensor has the form $T_{\mu\nu}^{\phi}e_{a}^{\mu}%
e_{b}^{\nu}=diag(\rho,p_{1},p_{2},p_{2},p_{2})$ where%

\begin{equation}
\rho+p_{1}=\frac{3\left(  \nu^{2}-1\right)  f(x)}%
{4\Omega(x)\eta^{2}}\geq0,\qquad\rho+p_{2}=0.%
\end{equation}

\section{Thermodynamics}

\label{thermo}

To gain some intuition, let us first consider the limit for which we can
recover the usual no-hairy planar black hole. This is possible in the limit
$\nu=1$ when the conformal factor becomes $\Omega(x)=\left[  \eta(x-1)\right]
^{-2}$. To obtain the usual planar coordinates, we should match the factors of
$d\Sigma^{2}$ so that $\Omega(x)=r^{2}$ and so we get (for the positive branch)%

\begin{equation}
x=1+\frac{1}{\eta r}%
\end{equation}
and for $\nu=1$ we obtain
\begin{equation}
f\Omega=r^{2}\left[  -\frac{\Lambda}{6}+\left(  \frac{1}{3}\right)  \left(
\frac{\alpha}{96\eta^{4}}\right)  \frac{1}{r^{4}}\right]  =\frac{r^{2}}{l^{2}%
}+\frac{1}{3}\left(  \frac{\alpha}{96\eta^{4}}\right)  \frac{1}{r^{4}}%
\end{equation}
The mass of the planar black hole was computed in \cite{Emparan:1999pm} by
using counterterms :
\begin{equation}
M=\frac{3}{16\pi G_{N}}mV_{3}=-\left(  \frac{1}{16\pi G_{N}}\right)
\frac{\alpha}{96\eta^{4}}V_{3}%
\end{equation}
where $m$ is the mass parameter of the planar black hole and $V_{3}$ is the
infinite volume of the space with the metric $d\Sigma^{2}$. It is worth
noticing that the event horizon exists only when $\alpha$ is negative, and
therefore the mass is always positive (as is also the temperature, see below).

Since in the planar case there is no Casimir energy, it is more convenient
when the scalars are turned on to use the method of Ashtekar, Das, and Magnon
\cite{Ashtekar:1999jx}. Interestingly enough, the mass has the same form as
for the planar black hole discussed above:
\begin{equation}
M_{hairy}=-\left(  \frac{1}{16\pi G_{N}}\right)  \frac{\alpha}{96\eta^{4}%
}V_{3}%
\end{equation}
but now the parameter $\alpha$ plays the role of a multiplicative coupling
constant in the Lagrangian and can not be eliminated as in the previous case,
when just the mass parameter characterizes the solution.

To check the first law, we also need the temperature and entropy of the hairy
black hole that are
\begin{equation}
T=\frac{f^{\prime}(x)}{4\pi\eta}\Big|_{x=x_{+}}=-\frac{ \alpha\left\vert
x_{+}^{\nu}-1\right\vert ^{3}}{288\pi\eta\nu^{3}x_{+}^{\frac{3}{2}\left(
\nu-1\right)  }}%
\end{equation}
\begin{equation}
S=\frac{A}{4G_{N}}=\frac{\Omega(x_{+})^{\frac{3}{2}}}{4G_{N}}V_{3}=\frac
{\nu^{3}x_{+}^{\frac{3}{2}\left(  \nu-1\right)  }}{4G_{N}\,\eta^{3}\left\vert
x_{+}^{\nu}-1\right\vert ^{3}}V_{3}%
\end{equation}
where $x_{+}$ is the location of the horizon, which is the largest root of
$f(x_{+})=0$.

Now, it is easy to check that the first law is satisfied if we work with
densities of the relevant physical quantities. It is well known that, with the
Wick rotation $t\rightarrow i\tau$, the Euclidean path integral yields a
thermal partition function. Then, the Euclidean black hole has the
interpretation of a saddle-point in this path integral and so the gravity
action evaluated for the classical solution is the leading contribution to the
free energy:
\begin{equation}
F=I_{E}T=M-TS=\frac{\alpha}{3^{2}2^{9}\pi G_{N}\eta^{4}}V_{3}<0.
\end{equation}
At this end, we would like to also comment on the physics of planar black holes 
when identifications in the horizon geometry are made so that the horizon becomes 
a torus.\footnote{When considering black holes with hyperbolic horizon topology, by doing 
identifications in the horizon geometry one can obtain topological black holes whose horizon 
topology is a generalization of the Riemann surfaces.} In this case, since 
there exist compact dimensions, there is a (negative) 
Casimir contribution to the black hole mass. Since this is constant, it does not play any role 
for the first law. Also, the quantum statistical relation is satisfied because the action gets 
also a contribution that will cancel the Casimir contribution from the mass.

However, the phase diagram is  drastically changed in this case. As in the no-hairy planar 
black hole case \cite{Surya:2001vj}, we found \cite{noi} that there is a similar phase transition 
between the hairy black holes and a hairy AdS soliton that is obtained by a double analytic
continuation (as in \cite{Horowitz:1998ha}) from the planar hairy black hole.

\section{Holography}

In the AdS/CFT context, the CFT on the boundary is the UV fixed point of a
$4$-dimensional QFT. Deformation of the gauge theory by the addition of
relevant operators is one way to reduce its symmetries. Using the gravity side
of the correspondence (deformations of $AdS_{5}$), one can obtain holographic
RG flows \cite{Akhmedov:1998vf, Freedman:1999gp} corresponding to
non-conformal field theories. Here, we consider the decoupling limit of
AdS/CFT duality at finite temperature \cite{Maldacena:1997re} when the bulk
theory is a black hole:
\begin{equation}
ds^{2}=b^{2}(u)\left[  -f(u)dt^{2}+d\Sigma^{2}+\frac{du^{2}}{f(u)}\right]
\end{equation}
and the boundary where the field theory is at the UV fixed point is
$u\rightarrow0$.

The running of the gauge coupling is simply a consequence of the dilaton being
non-constant. The beta-function of the theory, as a function in terms of the
background solution, is
\begin{equation}
\beta(e^{\phi})=b(u)\frac{de^{\phi(u)}}{db(u)} \label{beta1}%
\end{equation}

Given this definition, it is straightforward to write the beta-function in the
coordinates of (\ref{Ansatz})%

\begin{equation}
\beta(e^{\phi}) = 2\Omega\frac{de^{\phi(x)}}{d\Omega}=2\frac{\Omega}
{\Omega^{\prime}}\frac{de^{\phi(x)}}{dx}=2\frac{\Omega}{\Omega^{\prime}}
\frac{dx^{l_{\nu}^{-1}}}{dx} =-\frac{2}{l_{\nu}}\frac{e^{\nu\phi l_{\nu}}-1}{
e^{\nu\phi l_{\nu}}\nu+e^{\nu\phi l_{\nu}}+\nu-1}e^{\phi} \label{beta2}%
\end{equation}
As expected, we can easily see that the $\beta$-function vanishes at the
conformal UV fixed point.

In general, the beta-function is derived from the moduli potential via a
superpotential. In our case, since we have exact solutions we can use directly
(\ref{beta1}). However, for completeness, let us also discuss the case
$\alpha=0$ for which the superpotential can be explicitly obtained. Unlike the
$\alpha\neq0$ case, for $\alpha=0$ we obtain a domain wall with a naked
singularity. The superpotential is
\begin{equation}
W(\phi)=\frac{\sqrt{6}}{2\nu l}\left[  \left(  \nu-1\right)  e^{-\frac
{(\nu+1)\phi l_{\nu}}{2}}+\left(  \nu+1\right)  e^{\frac{(\nu-1)\phi l_{\nu}%
}{2}}\right]
\end{equation}
where
\begin{equation}
V(\phi)=3\left[  \frac{dW(\phi)}{d\phi}\right]  ^{2}-2W(\phi)^{2}%
\end{equation}
We can then compute the $\beta$-function from the superpotential as in
\cite{verlinde} $\beta(\phi)=-\frac{1}{W(\phi)}\frac{dW(\phi)}{d\phi}$ and see
that, indeed, has the same expression as (\ref{beta2}) (the $\beta$-function
does not depend explicitly on $\alpha$).\footnote{In fact, since in
\cite{verlinde} the beta function is defined as $\beta=a\frac{d\phi}{da}$ and
not with respect to $e^{\phi}$, the results match up to an $e^{\phi}$ factor.}

Let us now compute the c-function that is an off-shell generalization of the
central charge. The central charge counts the number of massless degrees of
freedom in the CFT, in other words it counts the ways in which the energy can
be transmitted. The coarse graining of a quantum field theory removes the
information about the small scales and so, for a QFT RG flow, there should
exist a c-function that is decreasing monotonically from the UV regime (large
radii in the dual AdS space) to the IR regime (small radii in the gravity
dual) of the QFT.

Here, we follow closely \cite{Freedman:1999gp}. By using the null energy
condition (that holds in our case) and the equations of motion, one can show
that for an ansatz:
\begin{equation}
ds^{2}=-a(r)^{2}dt^{2}+\frac{dr^{2}}{c^{2}(r)}+b^{2}(r)d\Sigma^{2}%
\end{equation}
the c-function is $C(r)=C_{0}\frac{a^{3}}{b^{\prime3}c^{3}}$, which in our
coordinates becomes
\begin{equation}
C(x)=8C_{0}\frac{\Omega^{\frac{9}{2}}}{\Omega^{\prime3}}=8C_{0}\frac{\nu^{3}%
}{\eta^{3}}\frac{x^{\frac{3}{2}(\nu+1)}}{(\nu x^{\nu}+\nu+x^{\nu}-1)^{3}}%
\end{equation}
and has the right properties. The constant is fixed by the entropy of black
hole (in the IR regime). For the planar neutral black hole, the conformal
radius is $b(r)=r$ and the c-function is constant, the flow is trivial (this
corresponds to `hairless' limit $\nu=1$).

We see that the c-function is completely determined by the conformal factor
and the other function in the metric, $f(x)$, does not play any role. We can
understand why since, in the dual description, the black hole is interpreted
as a thermal state. Therefore, the c-function should remain unchanged when we
excite a finite temperature vacuum in the same theory (with the same degrees
of freedom).

We would like to explain why the parameter $\alpha$ controls the physics in
deep infrared only. For this, let us expand the potential around the boundary
($\phi=0$):
\begin{align}
V(\phi)  &  =-\frac{12}{l^{2}}+\frac{\left(  \nu^{2}-1\right)  }{l^{2}}%
(-\frac{9l_{\nu}^{2}}{4}\phi^{2}+\frac{l_{\nu}^{3}}{4}\phi^{3}-\frac{3\left(
\nu^{2}-2\right)  l_{\nu}^{4}}{16}\phi^{4}+\frac{\left(  7\nu^{2}-18\right)
l_{\nu}^{5}}{80}\phi^{5}\\
&  +\frac{\left(  420-12\nu^{2}\left(  3\nu^{2}+5\right)  +5\alpha
l^{2}\right)  l_{\nu}^{6}}{5760}\phi^{6}+...)\nonumber
\end{align}
where the first non-trivial contribution of the $\alpha$ term was included.

Interestingly, it was shown in \cite{Henneaux:2006hk} that the $O(\phi
^{3})+...$ terms in the scalar potential are irrelevant to calculating the
Hamiltonian generators. Since for our solutions $\alpha$ appears at the $6$th
order in the expansion, this part of the scalar field potential does not
backreact strongly in the boundary, but in the bulk the self-interaction term
controlled by $\alpha$ is responsible for the appearance of the horizon.

We change the coordinates
\begin{equation}
\ln x=\frac{1}{\eta r}-\frac{1}{2\eta^{2}r^{2}}-\frac{\nu^{2}-9}{24\eta
^{3}r^{3}}+\frac{\nu^{2}-4}{12\eta^{4}r^{4}}+\frac{(9\nu^{2}-25)(\nu^{2}%
-25)}{1920\eta^{5}r^{5}} \label{form}%
\end{equation}
so that our metric matches the asymptotic form of \cite{Henneaux:2006hk}.

Solving the linearized scalar field equations with constant mass term, there
are two independent solutions. Since for our solutions $m^{2}l^{2}=-3$, we
obtain $\Delta_{+}=3$, $\Delta_{-}=1$ and so the behaviour of this scalar at
the boundary is%

\begin{equation}
\phi=\frac{a}{r}(1+...)+\frac{b}{r^{3}}(1+...)
\end{equation}

Depending on the form of the potential, there is a logarithmic term that can
appear when $\Delta_{+}/\Delta_{-}$ is an integer \cite{Henneaux:2006hk} (see,
also, \cite{Banados:2006de}), however it is not present in this solution.

(Super)gravity fields do not scale under $4$-dimensional conformal
transformations, so $a$ must have dimension $1$ and $b$ must have dimension
$3$ that corresponds to $a$ being a source (i.e. a mass) and $b$ being its VEV
(or condensate). Since $m^{2}<0$, the dual operators are relevant and so the
backreaction will be so that the metric remains asymptotically AdS, reflecting
the conformal fixed point in the UV. One can check this observation in the
coordinates of \cite{Henneaux:2006hk} for which we obtain the deviations from
the AdS metric at infinity as follows: the change of coordinates (\ref{form})
brings the metric to standard AdS coordinates. Indeed, the metric takes the form%

\begin{equation}
ds^{2}=-\frac{r^{2}}{l^{2}}dt^{2}+\frac{l^{2}}{r^{2}}dr^{2}+r^{2}\delta
_{mn}dx^{m}dx^{n}+h_{\mu\nu}dx^{\mu}dx^{\nu}%
\end{equation}
where $h_{\mu\nu}$ are the deviations from AdS. Using (\ref{form}) one can
then obtain that the fall off of the metric is the one predicted in
\cite{Henneaux:2006hk} for the generic case. We would like to point out that
the Ashtekar-Das-Magnon mass is exactly the $r^{-2}$ factor in the lapse function.

\section{Discussion and future directions}

Finding exact solutions of Einstein field equations with (or without) matter
sources is a subject of long standing interest. Indeed, exact solutions with
non-trivial moduli potentials can be important, in particular, for
phenomenological bottom-up approaches in string theory (see, e.g.,
\cite{gubser, kiritsis}). Some predictions of the gauge/gravity correspondence
may be universal enough as to apply to QCD, at least in certain regimes.

One important result we have obtained is the existence of a non-trivial RG
flow and, for future work, it will be interesting to understand how much the
dual field theory `mimics' QCD as in \cite{gubser, kiritsis}. In particular,
one could compute the speed of sound and bulk viscosity and see which values
of the `hairy parameter' ($\nu$) are interesting. There is also a conjecture
of Buchel on a dynamical bound of the bulk viscosity \cite{Buchel:2007mf} that
can be checked with our solutions.

Let us comment now on the limit $\alpha=0$ that is equivalent with considering
a vanishing mass for our solution. We see that the potential is still
non-trivial, but in this case we obtain a naked singularity. Once the
self-interaction of the scalar field, which is proportional to the parameter
$\alpha$ is considered the singularity gets `dressed' by a horizon and a
regular black hole solution is obtained. This is one concrete example for
which new terms/corrections of the potential convert singular solutions into
regular black holes with finite horizon area.

One possible extension of our work is to include gauge fields. The physics and
phase structure are richer in this case and it is possible to investigate the
physics at zero temperature due to the existence of extremal black hole
solutions. When the scalar field is non-minimally coupled to the gauge fields
and the scalar charge is determined by the charges and mass, the hair is
referred to as `secondary'. The scalar charge is not protected by a gauge
symmetry and so it is not a conserved charge. It was shown in
\cite{Gibbons:1996af} that the scalar charges, even if not conserved, can
appear in the first law of thermodynamics (the interpretation of this result,
though, should be taken with caution \cite{Astefanesei:2006sy}).

In the presence of gauge fields, we were also able to construct exact
solutions in AdS \cite{noi1}.\footnote{The existence of solutions for which
both modes of the scalar are normalizable was numerically proven in
\cite{Astefanesei:2008wz}.} In the extremal limit, since the flow of the
moduli is interpreted as an RG flow, the attractor mechanism acts as a no-hair
theorem \cite{Astefanesei:2007vh}.

There are other interesting directions, which can be investigated. For
example, using the results in Section \ref{thermo}, it is straightforward to
obtain the phase structure of our solutions. When the foliation of Euclidean
AdS is $R_{3}\times S^{1}$ ($S^{1}$ is the Euclidean time circle), the only
scale in the system (the temperature) can be scaled out via conformal
invariance. Thus, the $\mathcal{N}=4$ theory on $R_{3}$ cannot have a phase
transition at any nonzero temperature \cite{Witten:1998zw}. However, there are
first order transitions between these black holes and the corresponding
soliton (that is the solution with the minimum energy within the solutions
with the same boundary conditions), which is constructed by a double analytic
continuation similar with the one in \cite{Horowitz:1998ha}. It is important
to emphasize that there also are second order phase transitions similar to the
ones discussed in \cite{Martinez:2004nb}. These results will be presented in a
companion paper \cite{noi}.

To end our discussion we would like to point out that with our method we can
also generate exact solutions with scalar and gauge fields when the
cosmological constant is positive. These solutions can be described along the
lines of \cite{Leblond:2002ns} (e.g., computing the corresponding c-functions).

\section{Acknowledgments}

We would like to thank Guillaume Bossard, Arturo Gomez, Olivera Miskovic, Per
Sundell, and Stefan Theisen for helpful discussions and correspondence. DA
would also like to thank Nabamita Banerjee and Suvankar Dutta for interesting
correspondence. Research of A.A. is supported in part by the CONICYT anillo
grant ACT-91``Southern Theoretical Physics Laboratory''(STPLab) and by the
FONDECYT grant 11121187. The work of DA is supported by the Fondecyt Grant 1120446.

\end{document}